\documentclass{aa}  

\usepackage{graphicx}
\usepackage{txfonts}
\usepackage{natbib}
\begin{document}

   \title{DES J024008.08-551047.5: A new member of the polar ring galaxy family}
   \author{Akhil Krishna R \thanks{akhil.r@res.christuniversity.in}, Sreeja S Kartha, Blesson Mathew, K. Ujjwal, Savithri H. Ezhikode, Robin Thomas }

   \institute{Department of Physics and Electronics, CHRIST (Deemed to be University), Bangalore, India- 560029\\
              \email{akhil.r@res.christuniversity.in }}

    \authorrunning{Akhil et al. }
    \titlerunning{Discovery of a new polar ring galaxy }
 

  \abstract
   {}
   {This study presents the discovery of a new polar ring galaxy (PRG) candidate and highlights its unique features and characteristics. We provide evidence from photometric analysis that supports the inclusion of galaxy DES J024008.08-551047.5 (DJ0240) in the PRG catalogue.}
   {During the visual observations of optical imaging data obtained from the Dark Energy Camera Legacy Survey, a serendipitous discovery was made of a ringed galaxy, DJ0240. We conducted a one-dimensional isophotal analysis to determine the position angle of the ring component and its relative orientation to the host galaxy. A two-dimensional GALFIT analysis was performed to confirm the orthogonal nature of the ring galaxy and identify distinct components within the host galaxy. We compared the photometric properties of the host and ring components of DJ0240 with PRGs and other ring-type galaxies (RTGs), finding that DJ0240 shares similar properties with both of these galaxy types.}
   {We have discovered the galaxy DJ0240, a PRG candidate with a ring component positioned almost perpendicular to the host galaxy. The position angles of the ring and host components are $\sim$80 and $\sim$10 degrees, respectively, indicating that they are nearly orthogonal to each other. The extension of the ring component is three times greater than that of the host galaxy and shows a distinct colour separation, being bluer than the host. The estimated {\it g - r} colour values of the host and ring components are 0.86$\pm$0.02 and 0.59$\pm$0.10 mag, respectively. The colour value of the ring component is similar to those of typical spiral galaxies. The host galaxy's colour and the presence of a bulge and disc components indicate that the host galaxy may be lenticular. Our findings reveal a subtle yet noticeable colour difference between the host and ring components of PRGs and RTGs. We observe that both the host and ring components of DJ0240 align more closely with PRGs than with RTGs. Furthermore, we compared the Sersic index values of the ring component (n$_{ring}$) of galaxy DJ0240 with a selected sample of PRGs and Hoag-type galaxies. The results show that DJ0240 has a remarkably low n$_{ring}$ value of 0.13, supporting the galaxy's classification as a PRG. Hence, we suggest that the ring galaxy DJ0240 is a highly promising candidate for inclusion in the family of PRGs. }
   {}

   \maketitle

\section{Introduction}
During the visual examination of Dark Energy Camera Legacy Survey \citep[DECaLS;][]{2019Dey} data of the southern galaxy NGC 1031, a serendipitous discovery was made: a previously unknown ringed galaxy located  $\sim$ $36\arcmin $ south-east of NGC 1031. It has a distinctive combination of morphological and photometric characteristics, suggesting it might be classified as a polar ring galaxy (PRG).

Polar ring galaxies are a rare class of galaxies composed of two distinct components: a host galaxy and a ring that orbit in a nearly polar or orthogonal plane \citep{whitmore1990}. The central spheroid component is morphologically an E/S0 type galaxy with an old stellar population \citep{2015Reshetnikov}, whereas the ring component contains young stars {\citep{2002Iodice}}. The proposed mechanisms for the formation of PRGs are galaxy mergers \citep{1998kenjibekki,2003Bournaudcombes}, accretion from other galaxies \citep{reshkinov1997,2003Bournaudcombes}, and cold accretion from intergalactic medium filaments \citep{2006macci,2008Brook}. Moreover, the morphology of the polar structures varies widely and can take the form of a narrow ring, a wide annulus, a spindle, or even an inner polar structure \citep{2014Iodice}.

Despite the distinctive morphological and physical characteristics that PRGs exhibit, they remain highly uncommon. About 400 possible PRGs have been identified to date \citep{whitmore1990,moiseev2011MNRAS.418..244M,RM2019MNRAS.483.1470R}.  However, spectroscopic observations have only confirmed a  few dozen PRGs {\citep{whitmore1990,2011Reshetnikov_kin_conf}}, as the confirmation process is challenging. Utilising photometry to identify PRG candidates, we can streamline the selection process for spectroscopic observations of PRGs and enhance the confirmation rate. Detecting and confirming new PRGs is valuable as they offer unique opportunities to study the dynamics of galaxies, understand the formation processes of peculiar structures \citep{ordernes2016A&A...585A.156O}, investigate the effects of galaxy interactions {\citep{2004Combes_prg-inter}}, and probe the nature of dark matter \citep{2013Lghausenkroupa,2014Khoperskov}. These peculiar objects are easily identifiable when their central component and polar ring are observed edge-on \citep{2014Iodice,2022Nishimura}. The process of recognising a face-on ringed galaxy as a PRG involves the utilisation of various scientific methods. \cite{whitmore1990} developed a classification scheme to identify PRG candidates based on their morphology, kinematics, and orientation. Their catalogue consists of a total of 157 objects. Among them, only six objects already had a kinematic confirmation, and 27 galaxies were included in the category of `good' candidates. In addition, \citet{moiseev2011MNRAS.418..244M} identified 70 galaxies as `best' PRG candidates using data obtained from the Sloan Digital Sky Survey. Also, recent studies \citep{2011Finkelman,2022Nishimura} have reported a few individual PRG detections. However, the number of PRGs listed as potential candidates based on photometry remains relatively small. 

In this study we present the discovery of a ringed galaxy (DES J024008.08-551047.5, hereafter DJ0240) that exhibits a set of morphological and photometric features that qualify it as a possible PRG. Details regarding the object and data are presented in Sect. \ref{sec:data}. Section \ref{sec:data_analysis} outlines the analysis conducted and presents the corresponding results. Finally, we discuss and summarise the results in Sect. \ref{sec:dis-sum}. 
\section{Data inventory}
\label{sec:data}
\begin{figure}
    \centering
    \includegraphics[width=0.85\columnwidth]{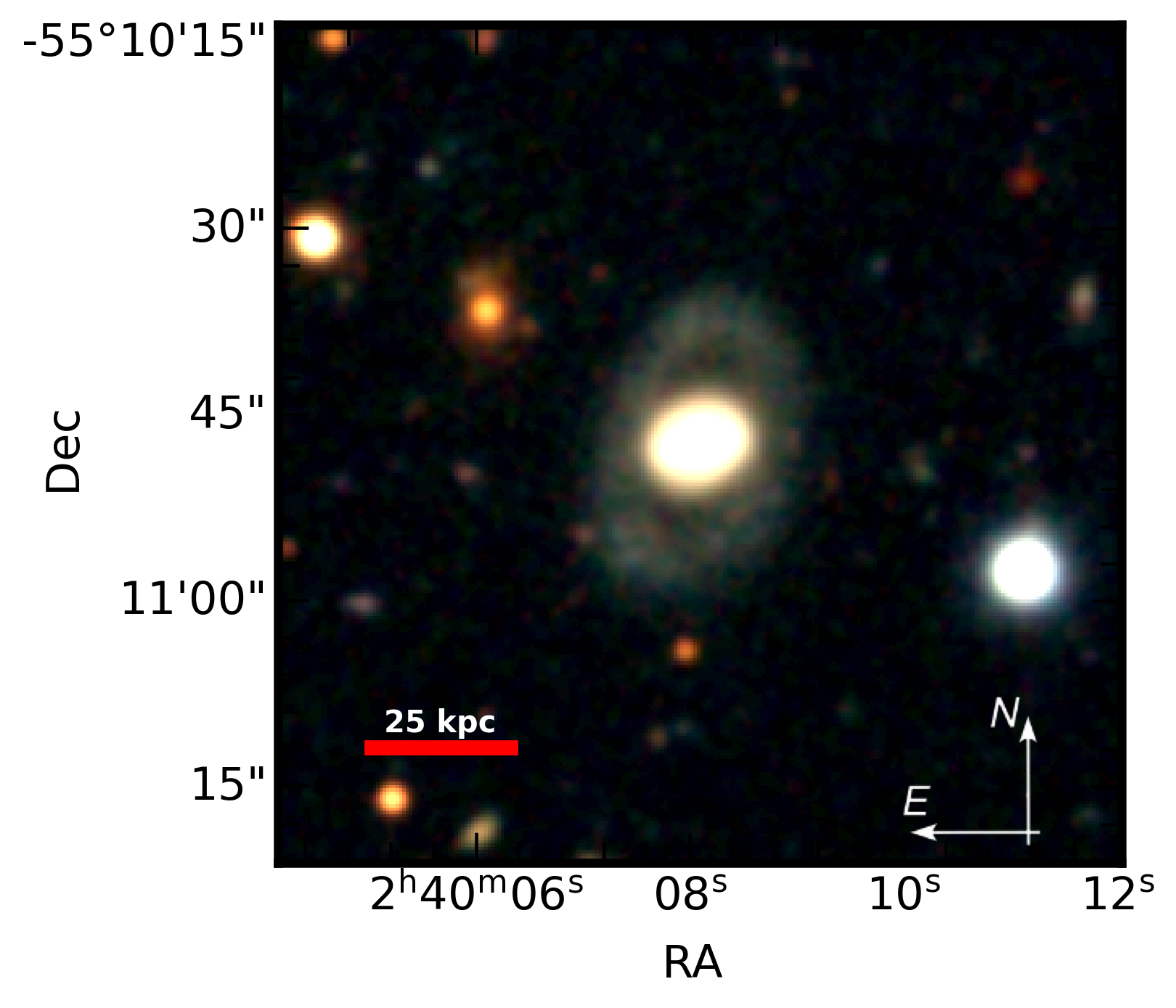}
    \caption{Optical colour composite image of the galaxy DJ0240. The DECaLS {\it g, r,} and {\it z} images are colour-coded in blue, green, and red, respectively, with a field of view of 1.125\arcmin x 1.125\arcmin. }
    \label{fig:color}
\end{figure}

We studied DJ0240, located at a photometric redshift of 0.12$\pm$0.02, using the south field (brick 0399m552) observation of DECaLS data release 10 \citep[DR10;][]{2019Dey}. DECaLS uses the Dark Energy Camera, which comprises 62 2k × 4k CCDs with a pixel scale of 0.262 arcsec/pixel and a 3.2 deg$^{2}$ field of view and is located at the 4 m Blanco telescope of Cerro Tololo Inter American Observatory \citep{2019Dey}. The Legacy Surveys DR10 imaging data is processed using the NOIRLab Community Pipeline \citep{2014Valdes} and the Legacypipe pipeline \citep{2016LangDustin}. The co-added {\it g-, r-,} and {\it z}-band images used in our analysis were calibrated in nanomaggies. The flux of one nanomaggie corresponds to an AB magnitude of 22.5 \citep{2016Driver,2019Dey}. The {\it g, r,} and {\it z} imaging data of DJ0240 were obtained from the DECaLS data archive. The colour composite of the galaxy DJ0240 is shown in Fig. \ref{fig:color}, and its basic properties are listed in Table \ref{tab:param}.\par
An extinction correction was applied to all the magnitudes of DJ0240. For the DECaLS filters, extinction coefficients computed by \cite{2011Schlafly} were utilised. To determine the extinction values for each filter, we multiplied the extinction coefficients by the E(B-V) values obtained from \citet{1998schlegel_SFD} at the given coordinates (RA = 40.0337, Dec = -55.1799). We obtain an E(B-V) value of 0.029 at the coordinates and extinction values of 0.09, 0.06, 0.04, and 0.03 for the {\it g, r, i,} and {\it z} filters, respectively. A flat Universe cosmology is adopted throughout this paper, with H$_{0}$ = 71 kms$^{-1} Mpc^{-1}$ and $\Omega$ = 0.27 \citep{2011Komatsu}.
\begin{table}
\caption{Basic properties of the galaxy DJ0240. References: ${^1}$\citet{20062mass}, ${^2}$\citet{2021Abbott_decals},  ${^3}$\citet{2022Duncan_decals}, ${^4}$\citet{2014Bilicki}}
\label{tab:param}
\scalebox{0.8}{
\begin{tabular}{ll}
\hline
{Parameter}   & {Value} \\ \hline
Name$^{1,2}$                 &DES J024008.08-551047.5 \\ 
RA (J2000)$^{1}$           & 40.0337       \\ 
Dec (J2000)$^{1}$          & -55.1799       \\ 
${\it g}_{mag}$ $^{2}$           & 17.7$\pm$0.002          \\ 
${\it r}_{mag}$ $^{2}$           & 16.8$\pm$0.001           \\ 
${\it i}_{mag}$ $^{2}$           & 16.2$\pm$0.001          \\ 
${\it z}_{mag}$ $^{2}$           & 16.1$\pm$0.002        \\ 
Photometric redshift$^{3,4}$       & 0.12$\pm$0.02          \\ 
Photometric distance (Mpc)       &  553$\pm$85           \\ 
{$M_{r}$ (mag)}                            & -21.9$\pm$0.3 \\
{Metric scale (kpc/ $^{''}$)}                       & 2.25  \\\hline
\end{tabular}}
\end{table}
\section{Analysis and results}
\label{sec:data_analysis}
\subsection{One-dimensional  surface photometry}
\begin{figure}
    \centering
    \includegraphics[width=0.9\columnwidth]{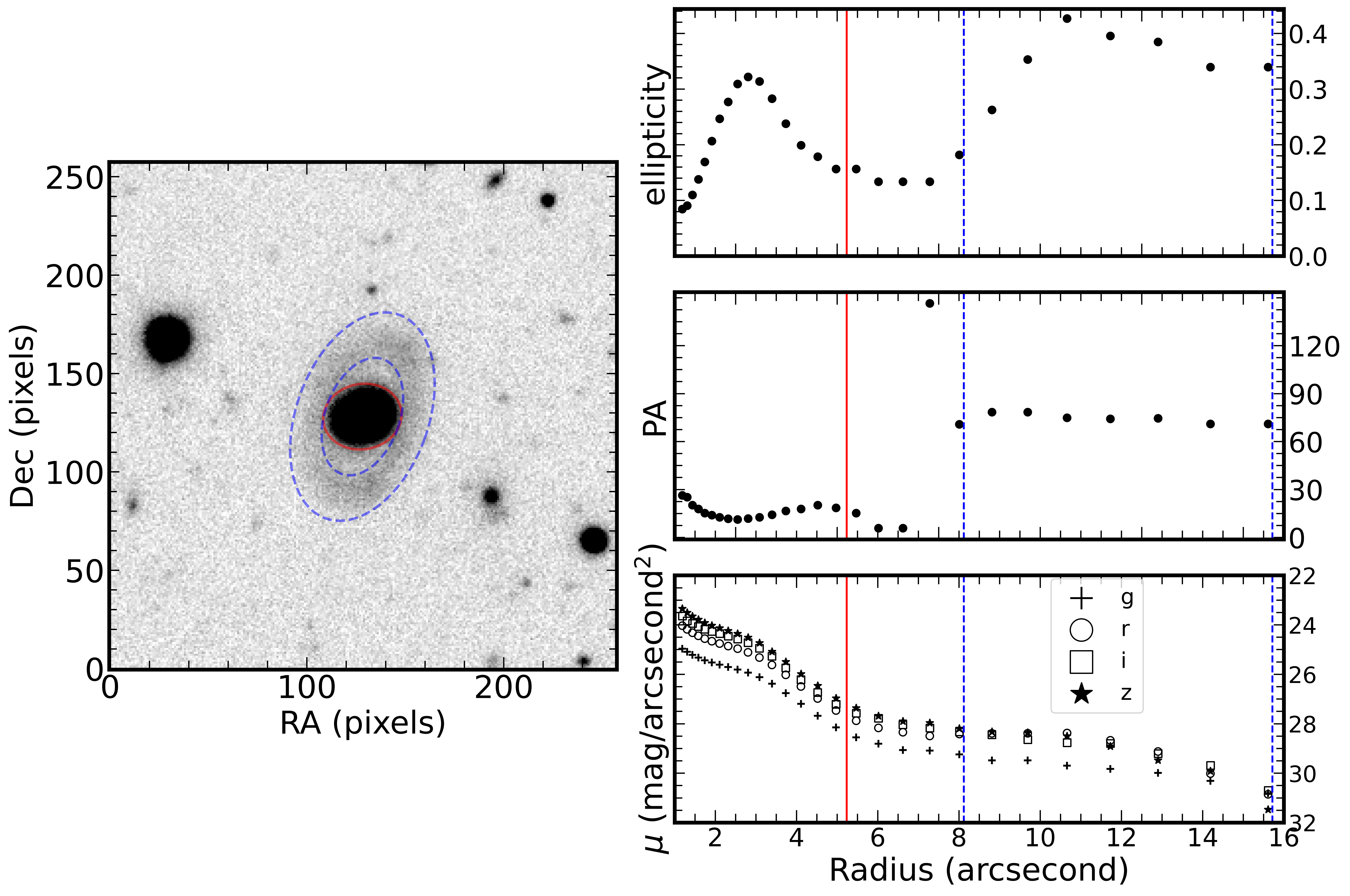}
    \caption{The left panel displays the separated ellipses for the host and ring components of a PRG overlaid on the r-band image. The right panel shows the corresponding ellipticity, position angle profiles, and surface brightness profiles of DJ0240. The solid red and dashed blue lines in both panels correspond to the extent of the host and ring components, respectively.}
    \label{fig:an_elip_pa}
\end{figure}

The isophotal analysis is an effective technique for examining various aspects of galaxies at different levels of surface brightness. It provides valuable insights into their structure, morphology, and composition \citep{2009Kormendy}. In this study, the surface photometry for the {\it r} filter was performed using Python's elliptical isophote analysis, as described by \cite{jedrzejwski1987}. The fitting process allowed the isophotes' centre, ellipticity, and position angle (measured anti-clockwise from the west) to vary freely. The fitted elliptical isophotes, shown in Fig. \ref{fig:an_elip_pa}, exhibit significant variations in their position angle and ellipticity at a radius of 20 pixels ($\sim$ 5\arcmin\arcmin) and 30 pixels ($\sim$ 8\arcmin\arcmin). Then it remains constant until 60 pixels ($\sim$ 15\arcmin\arcmin).  It is clear from the observations that the luminosity variations in the host galaxy and ring structure are responsible for these significant variations. We see a variation of $\sim$ 80 degrees in the position angle between the centre of the galaxy and the ring component, leading us to characterise the galaxy as a potential PRG candidate. The ellipticity and position angle distribution are illustrated in the right panel of Fig. \ref{fig:an_elip_pa}. We separated the ring component from the galaxy based on the observations shown in the left panel of Fig. \ref{fig:an_elip_pa}. The surface brightness profile is also included in the bottom-right panel of Fig. \ref{fig:an_elip_pa}; it reveals a brightness bump corresponding to the ring component after a radius of  $\sim$ 8\arcmin\arcmin up to $\sim$ 15\arcmin\arcmin.
The radius of the ring component is three times greater than that of the host component of DJ0240. The host component exhibits a position angle of $\sim$ 10 degrees, while the ring component displays a position angle of $\sim$ 80 degrees. This significant difference in position angles suggests that the host and ring components are nearly orthogonal. The host galaxy and the ring component have a similar ellipticity value of $\sim$ 0.3. It is important to note that the results presented here are based on the surface photometry analysis, which does not incorporate factors such as galaxy inclination and viewing angle. We estimated the colours of these components separately. The {\it g - r} values for the host and ring components are 0.86$\pm$0.02
 and 0.59$\pm$0.10 mag, respectively. The colour value of the ring component is similar to the average colour of spiral galaxies \citep{1995Fukugita,2007Fukugita,2013Tojeiro_color}. Also, the colour of the host galaxy suggests it is an early-type galaxy \citep{2007Fukugita,2010Nair_color,2013Tojeiro_color}.
Table \ref{tab:res} contains specific details regarding the photometry values we obtained. Based on the colour difference between the host component and the ring component, it is evident that the host component is significantly redder than the ring component.
\begin{table}
\caption{Properties of the galaxy DJ0240 obtained from this study. RE and PA correspond to the radial extent and position angle, respectively. All the magnitudes listed are corrected for extinction.}
\label{tab:res}
\scalebox{0.8}{
\begin{tabular}{llll}
\hline
{Parameters} & {Host} & {Ring} & {Total} \\ \hline
RE (\arcmin\arcmin)               & 5            & 15            & 15                    \\
PA (deg)                        &10             &80             &  -- \\
${\it g}$ (mag)                & 17.82$\pm$0.01  & 19.48$\pm$0.08 & 17.56$\pm$0.01           \\ 
${\it r}$ (mag)                & 16.971$\pm$0.006  & 18.89$\pm$0.05 & 16.759$\pm$0.002          \\ 
${\it i}$ (mag)                & 16.571$\pm$0.004  & 18.65$\pm$0.04  & 16.383$\pm$0.003          \\ 
${\it z}$ (mag)                & 16.327$\pm$0.003  & 18.49$\pm$0.03  & 16.150$\pm$0.002                \\
${\it g - r}$  (mag)              & 0.86$\pm$0.02           & 0.59$\pm$0.10            & 0.80$\pm$0.01                  \\
${\it g - i}$ (mag)               & 1.25$\pm$0.01           & 0.84$\pm$0.10           & 1.18$\pm$0.01                   \\ \hline
\end{tabular}}
\end{table}
\subsection{Two-dimensional  surface photometry}
\begin{figure}
    \centering
    \includegraphics[width=0.9\columnwidth]{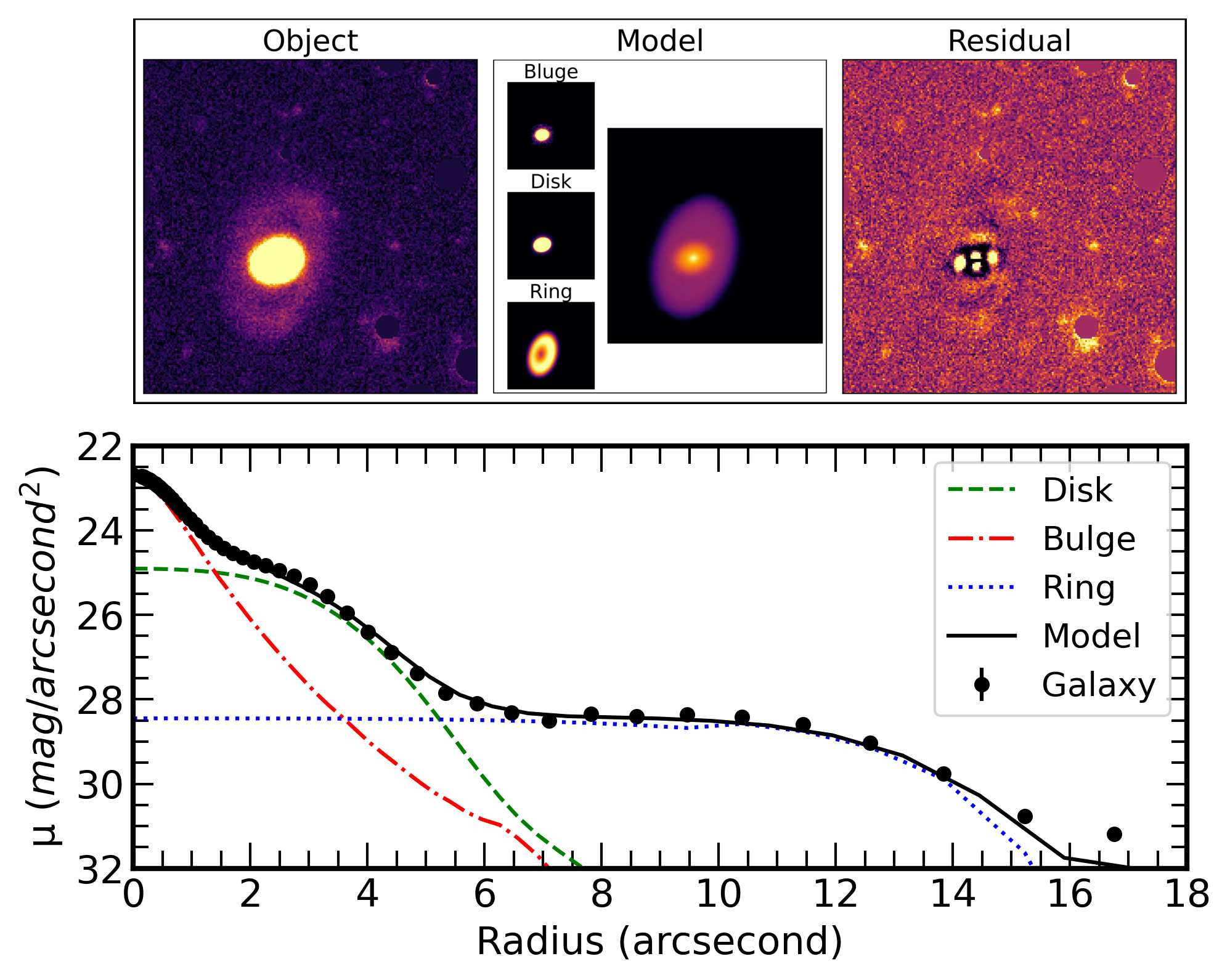}
    \caption{In the top image, the left panel showcases the DECaLS {\it r}-band image of the DJ0240. The centre panel displays the subcomponents and the total model obtained from the modelling process. The right panel exhibits the residual image. The bottom image illustrates the surface brightness profile for the  DECaLS {\it r}-band image. Various coloured lines are used to represent the subcomponents. Specifically, the dotted blue line corresponds to the truncated Sersic function, which represents the ring component.}
    \label{fig:galfit_out}
\end{figure}

We also performed parametric two-dimensional surface photometry fits with GALFIT \citep{2002Peng,2010peng}. GALFIT provides the flexibility to simultaneously model multiple parametric functions (such as Sersic, Gaussian, etc.). These functions can be modelled as multiple subcomponents of a single object, multiple objects within a frame, or a combination of the two. We used the DECaLS {\it r}-band image to do the photometry. To ensure accurate modelling, we masked all stars, background objects, and foreground objects in the images. Nearby unsaturated stars were used to estimate the point spread function at the object positions. A two-component model consisting of a host galaxy and a ring was used to fit the galaxy. We used two  Sersic functions to study the host galaxy and an inner truncated  Sersic function for the ring component \citep{2015Reshetnikov}. Both components were modelled with seven free parameters: object centre, total magnitude, effective radius, Sersic index, ellipticity, and position angle. Figure \ref{fig:galfit_out} displays the results obtained using GALFIT. The upper panel of the figure shows the input, model, and residual images from left to right. The bottom panel represents the total galaxy's brightness profiles and subcomponents. The results obtained for DJ0240 using GALFIT are listed in Table \ref{tab:galout}. Here, we observe two distinct components within the host galaxy -- a bulge and a disc. Typically, spiral and lenticular galaxies exhibit a combination of a bulge and a disc \citep{2009BarwaySudhanshu,2009Oohama}. This indicates that the host galaxy of DJ0240 is either spiral or lenticular. As mentioned in Sect. 3.1, the colour values obtained for the host galaxy indicate it is an early-type galaxy. Therefore, it is reasonable to suggest that the host galaxy is lenticular. Moreover, we observe that the effective radius of the ring component is approximately three times larger than that of the host galaxy. Also, there is a 70-degree difference in the position angle between the host galaxy and the ring component.  All these results indicate that DJ0240 is a promising PRG candidate.
\begin{table}
\caption{Result parameters of the GALFIT analysis for the galaxy DJ0240. The position angle is measured by taking due north as the zero degree and west as -90 degrees.}
\label{tab:galout}
\scalebox{0.8}{
\begin{tabular}{llll}
\hline
{{Parameters}} &{Sersic 1}& {Sersic 2}& {Sersic 3}\\\cline{2-4} 
 & \textit{{Bulge}} & \textit{{Disc}} & \textit{{Ring}}  \\ \hline
Integrated magnitude (mag)                 & 17.9 & 17.6 & 18.6 \\
Effective radius (\arcmin\arcmin)  & 0.6  & 2.5 & 8.9 \\
Sersic Index                         & 2.2  & 0.3 & 0.13 \\
Axis ratio                          & 0.65  & 0.7  & 0.6 \\
Position Angle (deg )               & -77  & -73  & -16     \\ 
\hline
\end{tabular}}
\end{table}
\renewcommand{\footnotesize}{\fontsize{7pt}{7pt}\selectfont}
\subsection{Known versus unknown}
{The photometric evidence indicates that DJ0240 is a potential PRG candidate. However, the lack of a distinct edge-on view and viewing angle of the galaxy means we cannot rule out that it is a face-on ring galaxy, for example a Hoag-type galaxy \citep[e.g. UGC 4599;][]{2023silchenko}. Hoag-type galaxies are perfect ring galaxies whose central galaxy is surrounded by a ring with ongoing star formation \citep{1950Hoag,2013BroschHoag}.} \par
\begin{figure}
    \centering
    \includegraphics[width=0.7\columnwidth]{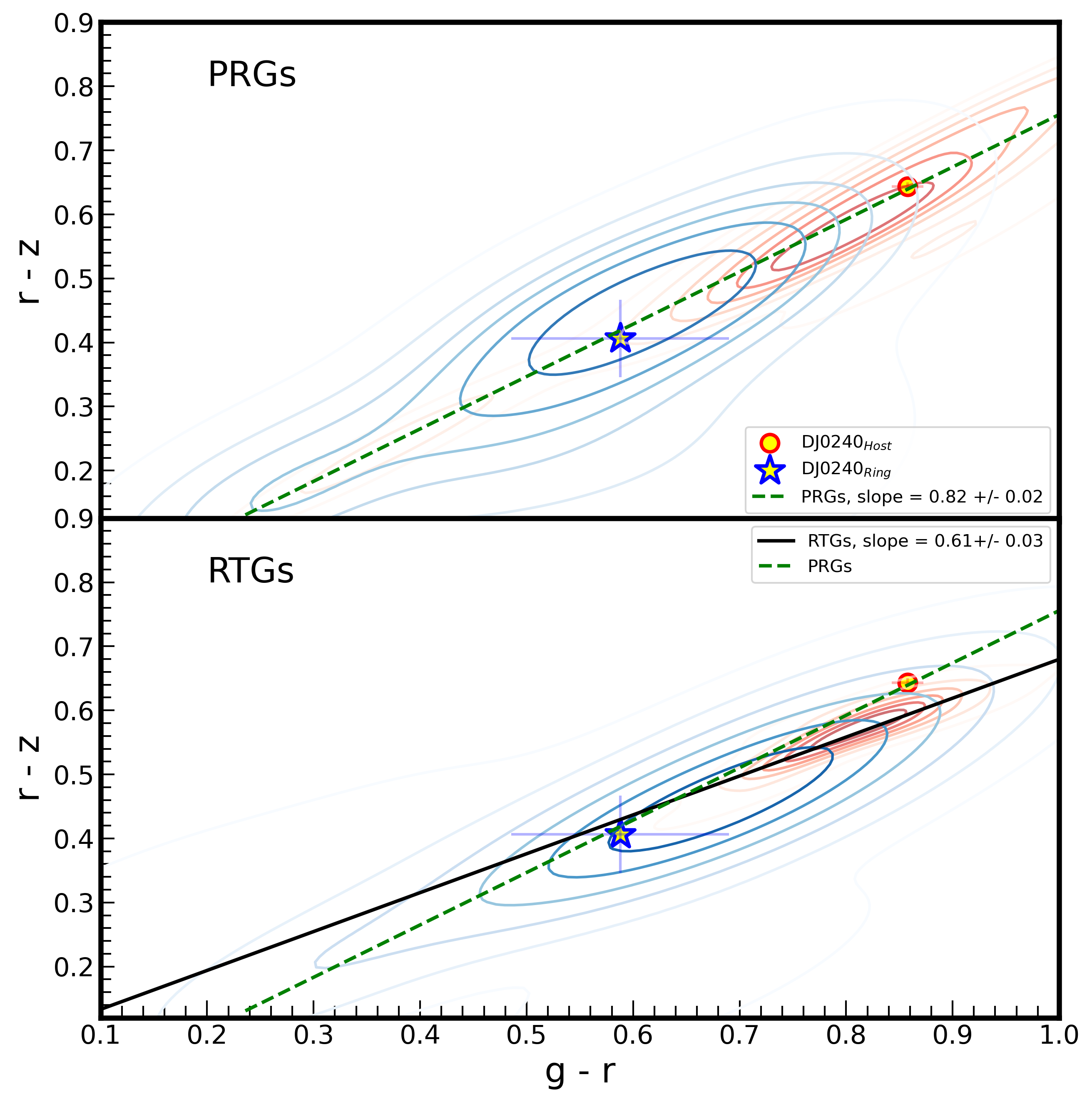}
    \caption{{ {\it g - r} vs {\it r - z} colour-colour diagram for the ring and host components of PRGs and RTGs, shown in the upper and lower panels, respectively. Blue contours denote the host components, and red contours represent the ring components of PRGs and RTGs. Additionally, the host component of DJ0240 is marked with a circle, and its ring component is indicated with a star symbol. The dashed green and solid black lines represent the best-fit line for PRGs and RTGs. The figure highlights a distinct difference in the slopes of the linear fits between PRGs and RTGs, and DJ0240 appears to align more closely with PRGs.}}
    \label{fig:CCD}
\end{figure}

To better constrain the classification of DJ0240, we conducted a comparative study with a selected sample of PRGs and other ring-type galaxies (RTGs). We selected the 70 best candidates of PRGs from the catalogue provided by \citet{moiseev2011MNRAS.418..244M}. We specifically selected the Hoag and elliptical families of rings, a total of 206 galaxies, from the catalogue presented by \citet{1998Faundez}. By utilising the {\it g-, r-,} and {\it z-} band data in DECaLS and based on the visibility, we narrowed our sample to 36 PRGs and 41 RTGs. We obtained the position angle and extent of the ring and host components for PRGs and RTGs from \citet{2013Smirnova} and \citet{1998Faundez}, respectively. We subsequently acquired the magnitudes for all these galaxies in the DECaLS optical filters {\it g, r,} and {\it z}\footnote{\footnotesize{The photometric data for all galaxies are available at https://github.com/akr777/Table}}. We then plotted a colour-colour diagram of the host and ring components of the PRGs, RTGs, and DJ0240, as shown in Fig. \ref{fig:CCD}. Our observations reveal a subtle but noticeable difference in the colour separation between the host and ring components of PRGs and RTGs. We estimate the difference in the {\it g - r} colour median value between RTGs' host and ring components to be 0.13 +/- 0.02, and 0.22 +/- 0.04 for  the PRGs. This suggests that PRGs tend to have a wider colour difference between their host and ring components, with a margin of 0.09 +/- 0.04 compared to RTGs. A linear relationship in the  {\it g - r} and {\it r - z} colour values is evident in both the host and ring components of PRGs and RTGs. However, there was a noticeable distinction in the slopes of the linear fits. In the case of PRGs, the linear relation exhibited a slope of 0.82 +/- 0.02, whereas a slope of 0.61 +/- 0.03 was identified for RTGs. Also, we observe that both the host and ring components of DJ0240 align more closely with those of the PRGs than those of the RTGs (as seen from Fig. \ref{fig:CCD}), providing further evidence for the classification of DJ0240 as a potential PRG. 
\begin{figure}
    \centering
    \includegraphics[width=0.7\columnwidth]{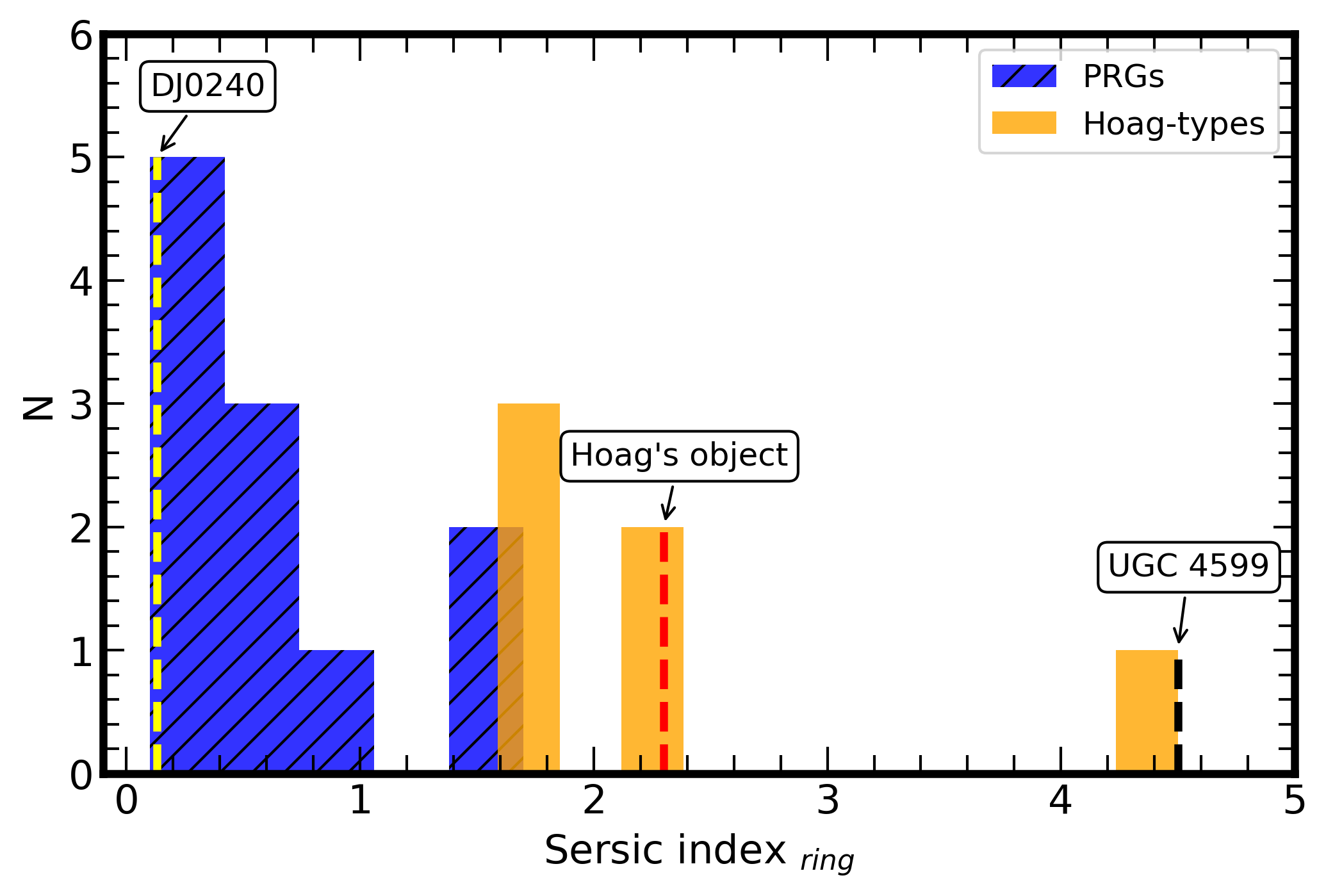}
    \caption{{Sersic index values of the ring components for ten PRGs, six Hoag-type galaxies, and DJ0240. The vertical dashed lines at n$_{ring}$ = 0.13, 2.3, and 4.5 correspond to DJ0240, Hoag's object, and UGC 4599, respectively.}}

    \label{fig:sersic}
\end{figure}

Although the host and ring components of DJ0240 exhibit a stronger alignment with PRGs than RTGs, the possibility of the object being a Hoag-type galaxy with an outer star-forming ring remains significant. \citet{RM2019MNRAS.483.1470R} and \citet{2022Mosenkov} note the low Sersic index values for the ring component (n$_{ring}$) of PRGs. Here, we aim to contrast the n$_{ring}$ value of DJ0240 with those of both PRGs and Hoag-type galaxies. However, it is important to note that the comprehensive comparison of n$_{ring}$ values between all PRGs and Hoag-type galaxies exceeds the scope of this study. So we selected the four Hoag-like galaxies from the RTG catalogue mentioned above. Also, we included two well-studied galaxies, Hoag's object \citep{1950Hoag} and  UGC4599 \citep{2011Finkelman_hoag,2023silchenko}. In addition to this, we obtained the n$_{ring}$ values of ten PRGs from \citet{2022Mosenkov}. Hence, our sample includes ten PRGs, six Hoag types, and DJ0240. Since the Hoag-type galaxy n$_{ring}$ values are not available in the literature, we modelled the ring component of these galaxies using an inner truncated Sersic function, as explained in Sect. 2. Figure \ref{fig:sersic} shows the n$_{ring}$ values of the PRGs, Hoag-type galaxies, and DJ0240. We note that the ring component of Hoag-type galaxies has n$_{ring}$ values greater than 1.6. Remarkably, our object, DJ0240, exhibited a very low Sersic index of n$_{ring}$ = 0.13, further evidence that it has a polar ring structure.  
\section{Discussion and summary}
\label{sec:dis-sum}
In the initial catalogue of PRGs, \citet{whitmore1990} established four categories of PRGs: best candidates, good candidates, possible candidates, and systems that may be related to PRGs. The best candidates are those that have been confirmed kinematically as PRGs. Good candidates must adhere to specific criteria, such as having a nearly orthogonal projected major axis for both components, nearly aligned centres for both components, and a luminous and nearly planar ring component comparable in size to the inner component. These factors are evaluated and discussed in the studies of \citet{2013Smirnova}, \citet{2015Reshetnikov}, and \cite{2022Mosenkov}, among others. In our study, we evaluated the structural components of DJ0240 to see whether it fits the good or best PRG candidate criteria of \citet{whitmore1990}.\par
Through one-dimensional isophotal analysis, we determined that the position angle of the ring component is $\sim$80 degrees and the position angle of the host galaxy is $\sim$10 degrees, which indicates that the ring component is nearly orthogonal to its host component. Additionally, by analysing the ellipticity and position angle profiles, we estimated the extent of both the host and ring components. Our analysis reveals that the ring component is three times more extended than the host galaxy. To further investigate the components of the galaxy, we conducted a two-dimensional GALFIT analysis. We were able to model the ring component of the galaxy using a truncated Sersic model. The GALFIT analysis also revealed that the ring component is nearly orthogonal to the host galaxy. We observed two distinct components within the host galaxy: a bulge and a disc. Also, we explored the Sersic indices and effective radius of each component, including the ring structure. The obtained results are comparable with those of other PRGs mentioned in the studies by \citet{2013Smirnova}, \citet{2015Reshetnikov}, and \citet{2022Mosenkov}. \par
Through the comparison of {\it g - r} and {\it r - z} colour values between the host and ring components of PRGs and RTGs, we identified a significant difference regarding their distributions of colours. We observe a distinct (albeit small) deviation in the slope of the colour distribution, which is intriguing since both types of galaxies share a common morphology. Furthermore, this observation supports the widely accepted definition of PRGs, that the ring component is typically bluer than the host component \citep{2015Reshetnikov,2022Mosenkov}. The presence of bluer ring components in PRGs indicates a higher level of ongoing star formation activity in these structures \citep{2017smirnov}. Also,  the PRGs are known to host early-type galaxies \citep{2015Reshetnikov}. From a sample of 36 PRGs and 41 RTGs, we observe that the ring component of PRGs and RTGs appears bluer than its central part. However, the colour separation between the host and ring components of PRGs is 0.09 +/- 0.04, wider when compared to RTGs (see Fig. \ref{fig:CCD}). Further research and investigations may help confirm and expand upon these intriguing results.  Despite this caveats, it is worth highlighting that our specific object, DJ0420, resembles PRGs more closely than RTGs. Also, the colour value of the ring structure exhibits a similarity to that of spiral galaxies \citep{2007Fukugita}. The colour value of the host galaxy and the presence of both bulge and disc components suggest that DJ0420 is a lenticular-type galaxy. 
Finally, \citet{RM2019MNRAS.483.1470R} note that the polar ring structure of PRGs can be modelled with a very low Sersic index (n$_{ring}$ < 0.1).  \citet{2022Mosenkov} discuss the presence of two groupings of polar structures: (i) true polar rings with n$_{ring}$  values less than 0.7 and (ii) polar structures with n$_{ring}$ > 1, which could possibly be polar discs. This study enabled us to accurately describe the ring of DJ0240 by utilising a Sersic function with an n$_{ring}$ of 0.13, shedding light on the ring structure; for comparison, the ring component of Hoag-type galaxies has n$_{ring}$ values greater than 1.6. The comparison with RTGs, including Hoag-type galaxies, reduces the possibility of DJ0240 belonging to a typical ring galaxy group and increases its likeliness of being a PRG.

In summary, this study presents the serendipitous discovery of a galaxy that exhibits a ring structure and has the potential to be classified as a PRG candidate. We utilised photometry on DECaLS optical images to explore the characteristics of the galaxy. The position angle and the colour distinction between the host and ring components are in line with the definition of PRGs. We have also discussed the possibility of the host galaxy being lenticular but suggest that the ringed galaxy DJ0240 is a highly promising candidate for inclusion in the PRG catalogue. However, further investigation is necessary to confirm this classification, particularly using spectroscopic observations to analyse the kinematic properties of both the host galaxy and the polar structure. 
\begin{acknowledgements}
We thank the anonymous referee for the valuable comments that improved the scientific content of the paper. AKR and SSK want to acknowledge the financial support from CHRIST (Deemed to be University, Bangalore) through the SEED money project (No: SMSS-2220, 12/2022 ). We thank our colleagues Ashish Devaraj, Arun Roy, and Sruthi K for their valuable comments on the manuscript.  We thank the Center for Research, CHRIST (Deemed to be University), for all their support during this work. AKR expresses sincere gratitude to Chien Y. Peng for his invaluable help and support during the GALFIT analysis. SSK and RT acknowledge the financial support from the Indian Space Research Organisation (ISRO) under the AstroSat archival data utilization program
(No. DS-2B-13013(2)/6/2019). UK acknowledges the Department of Science and Technology (DST) for the INSPIRE FELLOWSHIP (IF180855). This publication uses the data from the DECaLs survey. We gratefully thank all the individuals involved in the various teams for supporting the project from the early stages of the design to launch and observations with it in orbit. The complete acknowledgments for DESI Legacy Imaging Surveys can be found at https://www.legacysurvey.org/acknowledgment/.
\end{acknowledgements}
\bibliographystyle{aa}
\bibliography{example} 

\begin{thebibliography}{47}
\expandafter\ifx\csname natexlab\endcsname\relax\def\natexlab#1{#1}\fi

\bibitem[{{Abbott} {et~al.}(2021){Abbott}, {Adam{\'o}w}, {Aguena}, {Allam}, {Amon}, {Annis}, {Avila}, {Bacon}, {Banerji}, {Bechtol}, {Becker}, {Bernstein}, {Bertin}, {Bhargava}, {Bridle}, {Brooks}, {Burke}, {Carnero Rosell}, {Carrasco Kind}, {Carretero}, {Castander}, {Cawthon}, {Chang}, {Choi}, {Conselice}, {Costanzi}, {Crocce}, {da Costa}, {Davis}, {De Vicente}, {DeRose}, {Desai}, {Diehl}, {Dietrich}, {Drlica-Wagner}, {Eckert}, {Elvin-Poole}, {Everett}, {Evrard}, {Ferrero}, {Fert{\'e}}, {Flaugher}, {Fosalba}, {Friedel}, {Frieman}, {Garc{\'\i}a-Bellido}, {Gaztanaga}, {Gelman}, {Gerdes}, {Giannantonio}, {Gill}, {Gruen}, {Gruendl}, {Gschwend}, {Gutierrez}, {Hartley}, {Hinton}, {Hollowood}, {Honscheid}, {Huterer}, {James}, {Jeltema}, {Johnson}, {Kent}, {Kron}, {Kuehn}, {Kuropatkin}, {Lahav}, {Li}, {Lidman}, {Lin}, {MacCrann}, {Maia}, {Manning}, {Maloney}, {March}, {Marshall}, {Martini}, {Melchior}, {Menanteau}, {Miquel}, {Morgan}, {Myles}, {Neilsen}, {Ogando}, {Palmese}, {Paz-Chinch{\'o}n}, {Petravick},
  {Pieres}, {Plazas}, {Pond}, {Rodriguez-Monroy}, {Romer}, {Roodman}, {Rykoff}, {Sako}, {Sanchez}, {Santiago}, {Scarpine}, {Serrano}, {Sevilla-Noarbe}, {Smith}, {Smith}, {Soares-Santos}, {Suchyta}, {Swanson}, {Tarle}, {Thomas}, {To}, {Tremblay}, {Troxel}, {Tucker}, {Turner}, {Varga}, {Walker}, {Wechsler}, {Weller}, {Wester}, {Wilkinson}, {Yanny}, {Zhang}, {Nikutta}, {Fitzpatrick}, {Jacques}, {Scott}, {Olsen}, {Huang}, {Herrera}, {Juneau}, {Nidever}, {Weaver}, {Adean}, {Correia}, {de Freitas}, {Freitas}, {Singulani}, {Vila-Verde}, \& {Linea Science Server}}]{2021Abbott_decals}
{Abbott}, T.~M.~C., {Adam{\'o}w}, M., {Aguena}, M., {et~al.} 2021, \apjs, 255, 20

\bibitem[{{Barway} {et~al.}(2009){Barway}, {Wadadekar}, {Kembhavi}, \& {Mayya}}]{2009BarwaySudhanshu}
{Barway}, S., {Wadadekar}, Y., {Kembhavi}, A.~K., \& {Mayya}, Y.~D. 2009, \mnras, 394, 1991

\bibitem[{{Bekki}(1998)}]{1998kenjibekki}
{Bekki}, K. 1998, \apj, 496, 713

\bibitem[{{Bilicki} {et~al.}(2014){Bilicki}, {Jarrett}, {Peacock}, {Cluver}, \& {Steward}}]{2014Bilicki}
{Bilicki}, M., {Jarrett}, T.~H., {Peacock}, J.~A., {Cluver}, M.~E., \& {Steward}, L. 2014, \apjs, 210, 9

\bibitem[{{Bournaud} \& {Combes}(2003)}]{2003Bournaudcombes}
{Bournaud}, F. \& {Combes}, F. 2003, AAP, 401, 817

\bibitem[{{Brook} {et~al.}(2008){Brook}, {Governato}, {Quinn}, {Wadsley}, {Brooks}, {Willman}, {Stilp}, \& {Jonsson}}]{2008Brook}
{Brook}, C.~B., {Governato}, F., {Quinn}, T., {et~al.} 2008, \apj, 689, 678

\bibitem[{{Brosch} {et~al.}(2013){Brosch}, {Finkelman}, {Oosterloo}, {Jozsa}, \& {Moiseev}}]{2013BroschHoag}
{Brosch}, N., {Finkelman}, I., {Oosterloo}, T., {Jozsa}, G., \& {Moiseev}, A. 2013, \mnras, 435, 475

\bibitem[{{Combes}(2004)}]{2004Combes_prg-inter}
{Combes}, F. 2004, in Astrophysics and Space Science Library, Vol. 319, Penetrating Bars Through Masks of Cosmic Dust, ed. D.~L. {Block}, I.~{Puerari}, K.~C. {Freeman}, R.~{Groess}, \& E.~K. {Block}, 57

\bibitem[{{Dey} {et~al.}(2019){Dey}, {Schlegel}, {Lang}, {Blum}, {Burleigh}, {Fan}, {Findlay}, {Finkbeiner}, {Herrera}, {Juneau}, {Landriau}, {Levi}, {McGreer}, {Meisner}, {Myers}, {Moustakas}, {Nugent}, {Patej}, {Schlafly}, {Walker}, {Valdes}, {Weaver}, {Y{\`e}che}, {Zou}, {Zhou}, {Abareshi}, {Abbott}, {Abolfathi}, {Aguilera}, {Alam}, {Allen}, {Alvarez}, {Annis}, {Ansarinejad}, {Aubert}, {Beechert}, {Bell}, {BenZvi}, {Beutler}, {Bielby}, {Bolton}, {Brice{\~n}o}, {Buckley-Geer}, {Butler}, {Calamida}, {Carlberg}, {Carter}, {Casas}, {Castander}, {Choi}, {Comparat}, {Cukanovaite}, {Delubac}, {DeVries}, {Dey}, {Dhungana}, {Dickinson}, {Ding}, {Donaldson}, {Duan}, {Duckworth}, {Eftekharzadeh}, {Eisenstein}, {Etourneau}, {Fagrelius}, {Farihi}, {Fitzpatrick}, {Font-Ribera}, {Fulmer}, {G{\"a}nsicke}, {Gaztanaga}, {George}, {Gerdes}, {Gontcho}, {Gorgoni}, {Green}, {Guy}, {Harmer}, {Hernandez}, {Honscheid}, {Huang}, {James}, {Jannuzi}, {Jiang}, {Joyce}, {Karcher}, {Karkar}, {Kehoe}, {Kneib}, {Kueter-Young}, {Lan},
  {Lauer}, {Le Guillou}, {Le Van Suu}, {Lee}, {Lesser}, {Perreault Levasseur}, {Li}, {Mann}, {Marshall}, {Mart{\'\i}nez-V{\'a}zquez}, {Martini}, {du Mas des Bourboux}, {McManus}, {Meier}, {M{\'e}nard}, {Metcalfe}, {Mu{\~n}oz-Guti{\'e}rrez}, {Najita}, {Napier}, {Narayan}, {Newman}, {Nie}, {Nord}, {Norman}, {Olsen}, {Paat}, {Palanque-Delabrouille}, {Peng}, {Poppett}, {Poremba}, {Prakash}, {Rabinowitz}, {Raichoor}, {Rezaie}, {Robertson}, {Roe}, {Ross}, {Ross}, {Rudnick}, {Safonova}, {Saha}, {S{\'a}nchez}, {Savary}, {Schweiker}, {Scott}, {Seo}, {Shan}, {Silva}, {Slepian}, {Soto}, {Sprayberry}, {Staten}, {Stillman}, {Stupak}, {Summers}, {Sien Tie}, {Tirado}, {Vargas-Maga{\~n}a}, {Vivas}, {Wechsler}, {Williams}, {Yang}, {Yang}, {Yapici}, {Zaritsky}, {Zenteno}, {Zhang}, {Zhang}, {Zhou}, \& {Zhou}}]{2019Dey}
{Dey}, A., {Schlegel}, D.~J., {Lang}, D., {et~al.} 2019, \aj, 157, 168

\bibitem[{{Driver} {et~al.}(2016){Driver}, {Wright}, {Andrews}, {Davies}, {Kafle}, {Lange}, {Moffett}, {Mannering}, {Robotham}, {Vinsen}, {Alpaslan}, {Andrae}, {Baldry}, {Bauer}, {Bamford}, {Bland-Hawthorn}, {Bourne}, {Brough}, {Brown}, {Cluver}, {Croom}, {Colless}, {Conselice}, {da Cunha}, {De Propris}, {Drinkwater}, {Dunne}, {Eales}, {Edge}, {Frenk}, {Graham}, {Grootes}, {Holwerda}, {Hopkins}, {Ibar}, {van Kampen}, {Kelvin}, {Jarrett}, {Jones}, {Lara-Lopez}, {Liske}, {Lopez-Sanchez}, {Loveday}, {Maddox}, {Madore}, {Mahajan}, {Meyer}, {Norberg}, {Penny}, {Phillipps}, {Popescu}, {Tuffs}, {Peacock}, {Pimbblet}, {Prescott}, {Rowlands}, {Sansom}, {Seibert}, {Smith}, {Sutherland}, {Taylor}, {Valiante}, {Vazquez-Mata}, {Wang}, {Wilkins}, \& {Williams}}]{2016Driver}
{Driver}, S.~P., {Wright}, A.~H., {Andrews}, S.~K., {et~al.} 2016, \mnras, 455, 3911

\bibitem[{{Duncan}(2022)}]{2022Duncan_decals}
{Duncan}, K.~J. 2022, \mnras, 512, 3662

\bibitem[{{Faundez-Abans} \& {de Oliveira-Abans}(1998)}]{1998Faundez}
{Faundez-Abans}, M. \& {de Oliveira-Abans}, M. 1998, \aaps, 129, 357

\bibitem[{{Finkelman} \& {Brosch}(2011)}]{2011Finkelman_hoag}
{Finkelman}, I. \& {Brosch}, N. 2011, \mnras, 413, 2621

\bibitem[{{Finkelman} {et~al.}(2011){Finkelman}, {Graur}, \& {Brosch}}]{2011Finkelman}
{Finkelman}, I., {Graur}, O., \& {Brosch}, N. 2011, \mnras, 412, 208

\bibitem[{{Fukugita} {et~al.}(2007){Fukugita}, {Nakamura}, {Okamura}, {Yasuda}, {Barentine}, {Brinkmann}, {Gunn}, {Harvanek}, {Ichikawa}, {Lupton}, {Schneider}, {Strauss}, \& {York}}]{2007Fukugita}
{Fukugita}, M., {Nakamura}, O., {Okamura}, S., {et~al.} 2007, \aj, 134, 579

\bibitem[{{Fukugita} {et~al.}(1995){Fukugita}, {Shimasaku}, \& {Ichikawa}}]{1995Fukugita}
{Fukugita}, M., {Shimasaku}, K., \& {Ichikawa}, T. 1995, \pasp, 107, 945

\bibitem[{{Hoag}(1950)}]{1950Hoag}
{Hoag}, A.~A. 1950, \aj, 55, 170

\bibitem[{{Iodice}(2014)}]{2014Iodice}
{Iodice}, E. 2014, in Astronomical Society of the Pacific Conference Series, Vol. 486, Multi-Spin Galaxies, ed. E.~{Iodice} \& E.~M. {Corsini}, 39

\bibitem[{{Iodice} {et~al.}(2002){Iodice}, {Arnaboldi}, {Sparke}, \& {Freeman}}]{2002Iodice}
{Iodice}, E., {Arnaboldi}, M., {Sparke}, L.~S., \& {Freeman}, K.~C. 2002, \aap, 391, 117

\bibitem[{{Jedrzejewski}(1987)}]{jedrzejwski1987}
{Jedrzejewski}, R.~I. 1987, \mnras, 226, 747

\bibitem[{{Khoperskov} {et~al.}(2014){Khoperskov}, {Moiseev}, {Khoperskov}, \& {Saburova}}]{2014Khoperskov}
{Khoperskov}, S.~A., {Moiseev}, A.~V., {Khoperskov}, A.~V., \& {Saburova}, A.~S. 2014, \mnras, 441, 2650

\bibitem[{{Komatsu} {et~al.}(2011){Komatsu}, {Smith}, {Dunkley}, {Bennett}, {Gold}, {Hinshaw}, {Jarosik}, {Larson}, {Nolta}, {Page}, {Spergel}, {Halpern}, {Hill}, {Kogut}, {Limon}, {Meyer}, {Odegard}, {Tucker}, {Weiland}, {Wollack}, \& {Wright}}]{2011Komatsu}
{Komatsu}, E., {Smith}, K.~M., {Dunkley}, J., {et~al.} 2011, \apjs, 192, 18

\bibitem[{{Kormendy} {et~al.}(2009){Kormendy}, {Fisher}, {Cornell}, \& {Bender}}]{2009Kormendy}
{Kormendy}, J., {Fisher}, D.~B., {Cornell}, M.~E., \& {Bender}, R. 2009, \apjs, 182, 216

\bibitem[{{Lang} {et~al.}(2016){Lang}, {Hogg}, \& {Mykytyn}}]{2016LangDustin}
{Lang}, D., {Hogg}, D.~W., \& {Mykytyn}, D. 2016, {The Tractor: Probabilistic astronomical source detection and measurement}, Astrophysics Source Code Library, record ascl:1604.008

\bibitem[{{L{\"u}ghausen} {et~al.}(2013){L{\"u}ghausen}, {Famaey}, {Kroupa}, {Angus}, {Combes}, {Gentile}, {Tiret}, \& {Zhao}}]{2013Lghausenkroupa}
{L{\"u}ghausen}, F., {Famaey}, B., {Kroupa}, P., {et~al.} 2013, \mnras, 432, 2846

\bibitem[{{Macci{\`o}} {et~al.}(2006){Macci{\`o}}, {Moore}, \& {Stadel}}]{2006macci}
{Macci{\`o}}, A.~V., {Moore}, B., \& {Stadel}, J. 2006, \apjl, 636, L25

\bibitem[{{Moiseev} {et~al.}(2011){Moiseev}, {Smirnova}, {Smirnova}, \& {Reshetnikov}}]{moiseev2011MNRAS.418..244M}
{Moiseev}, A.~V., {Smirnova}, K.~I., {Smirnova}, A.~A., \& {Reshetnikov}, V.~P. 2011, MNRAS, 418, 244

\bibitem[{{Mosenkov} {et~al.}(2022){Mosenkov}, {Reshetnikov}, {Skryabina}, \& {Shakespear}}]{2022Mosenkov}
{Mosenkov}, A.~V., {Reshetnikov}, V.~P., {Skryabina}, M.~N., \& {Shakespear}, Z. 2022, Research in Astronomy and Astrophysics, 22, 115003

\bibitem[{{Nair} \& {Abraham}(2010)}]{2010Nair_color}
{Nair}, P.~B. \& {Abraham}, R.~G. 2010, \apjs, 186, 427

\bibitem[{{Nishimura} {et~al.}(2022){Nishimura}, {Matsubayashi}, {Murayama}, \& {Taniguchi}}]{2022Nishimura}
{Nishimura}, M., {Matsubayashi}, K., {Murayama}, T., \& {Taniguchi}, Y. 2022, \pasp, 134, 094105

\bibitem[{{Oohama} {et~al.}(2009){Oohama}, {Okamura}, {Fukugita}, {Yasuda}, \& {Nakamura}}]{2009Oohama}
{Oohama}, N., {Okamura}, S., {Fukugita}, M., {Yasuda}, N., \& {Nakamura}, O. 2009, \apj, 705, 245

\bibitem[{{Ordenes-Brice{\~n}o} {et~al.}(2016){Ordenes-Brice{\~n}o}, {Georgiev}, {Puzia}, {Goudfrooij}, \& {Arnaboldi}}]{ordernes2016A&A...585A.156O}
{Ordenes-Brice{\~n}o}, Y., {Georgiev}, I.~Y., {Puzia}, T.~H., {Goudfrooij}, P., \& {Arnaboldi}, M. 2016, AAP, 585, A156

\bibitem[{{Peng} {et~al.}(2002){Peng}, {Ho}, {Impey}, \& {Rix}}]{2002Peng}
{Peng}, C.~Y., {Ho}, L.~C., {Impey}, C.~D., \& {Rix}, H.-W. 2002, \aj, 124, 266

\bibitem[{{Peng} {et~al.}(2010){Peng}, {Ho}, {Impey}, \& {Rix}}]{2010peng}
{Peng}, C.~Y., {Ho}, L.~C., {Impey}, C.~D., \& {Rix}, H.-W. 2010, \aj, 139, 2097

\bibitem[{{Reshetnikov} \& {Combes}(2015)}]{2015Reshetnikov}
{Reshetnikov}, V. \& {Combes}, F. 2015, \mnras, 447, 2287

\bibitem[{{Reshetnikov} \& {Sotnikova}(1997)}]{reshkinov1997}
{Reshetnikov}, V. \& {Sotnikova}, N. 1997, AAP, 325, 933

\bibitem[{{Reshetnikov} {et~al.}(2011){Reshetnikov}, {Fa{\'u}ndez-Abans}, \& {de Oliveira-Abans}}]{2011Reshetnikov_kin_conf}
{Reshetnikov}, V.~P., {Fa{\'u}ndez-Abans}, M., \& {de Oliveira-Abans}, M. 2011, Astronomy Letters, 37, 171

\bibitem[{{Reshetnikov} \& {Mosenkov}(2019)}]{RM2019MNRAS.483.1470R}
{Reshetnikov}, V.~P. \& {Mosenkov}, A.~V. 2019, MNRAS, 483, 1470

\bibitem[{{Schlafly} \& {Finkbeiner}(2011)}]{2011Schlafly}
{Schlafly}, E.~F. \& {Finkbeiner}, D.~P. 2011, \apj, 737, 103

\bibitem[{{Schlegel} {et~al.}(1998){Schlegel}, {Finkbeiner}, \& {Davis}}]{1998schlegel_SFD}
{Schlegel}, D.~J., {Finkbeiner}, D.~P., \& {Davis}, M. 1998, \apj, 500, 525

\bibitem[{{Sil'chenko} {et~al.}(2023){Sil'chenko}, {Moiseev}, {Oparin}, {Beckman}, \& {Font}}]{2023silchenko}
{Sil'chenko}, O., {Moiseev}, A., {Oparin}, D., {Beckman}, J.~E., \& {Font}, J. 2023, \aap, 669, L10

\bibitem[{{Skrutskie} {et~al.}(2006){Skrutskie}, {Cutri}, {Stiening}, {Weinberg}, {Schneider}, {Carpenter}, {Beichman}, {Capps}, {Chester}, {Elias}, {Huchra}, {Liebert}, {Lonsdale}, {Monet}, {Price}, {Seitzer}, {Jarrett}, {Kirkpatrick}, {Gizis}, {Howard}, {Evans}, {Fowler}, {Fullmer}, {Hurt}, {Light}, {Kopan}, {Marsh}, {McCallon}, {Tam}, {Van Dyk}, \& {Wheelock}}]{20062mass}
{Skrutskie}, M.~F., {Cutri}, R.~M., {Stiening}, R., {et~al.} 2006, \aj, 131, 1163

\bibitem[{{Smirnova} \& {Moiseev}(2013)}]{2013Smirnova}
{Smirnova}, K.~I. \& {Moiseev}, A.~V. 2013, Astrophysical Bulletin, 68, 371

\bibitem[{{Smirnova} {et~al.}(2017){Smirnova}, {Wiebe}, \& {Moiseev}}]{2017smirnov}
{Smirnova}, K.~I., {Wiebe}, D.~S., \& {Moiseev}, A.~V. 2017, Open Astronomy, 26, 88

\bibitem[{{Tojeiro} {et~al.}(2013){Tojeiro}, {Masters}, {Richards}, {Percival}, {Bamford}, {Maraston}, {Nichol}, {Skibba}, \& {Thomas}}]{2013Tojeiro_color}
{Tojeiro}, R., {Masters}, K.~L., {Richards}, J., {et~al.} 2013, \mnras, 432, 359

\bibitem[{{Valdes} {et~al.}(2014){Valdes}, {Gruendl}, \& {DES Project}}]{2014Valdes}
{Valdes}, F., {Gruendl}, R., \& {DES Project}. 2014, in Astronomical Society of the Pacific Conference Series, Vol. 485, Astronomical Data Analysis Software and Systems XXIII, ed. N.~{Manset} \& P.~{Forshay}, 379

\bibitem[{{Whitmore} {et~al.}(1990){Whitmore}, {Lucas}, {McElroy}, {Steiman-Cameron}, {Sackett}, \& {Olling}}]{whitmore1990}
{Whitmore}, B.~C., {Lucas}, R.~A., {McElroy}, D.~B., {et~al.} 1990, AJ, 100, 1489

\end{thebibliography}
\end{document}